\documentclass[aps, prl, preprint, floatfix, nofootinbib, longbibliography]{revtex4-1}

\usepackage[english]{babel}

\usepackage[letterpaper,top=2cm,bottom=2cm,left=3cm,right=3cm,marginparwidth=1.75cm]{geometry}

\usepackage{amsmath}
\usepackage{graphicx}
\usepackage[colorlinks=true, allcolors=blue]{hyperref}
\usepackage{titlesec}
\titlespacing{\section}{0pc}{1pc}{1pc}
\titlespacing{\subsection}{3ex}{1pc}{1pc}
\usepackage{lineno}
\usepackage{soul}
\usepackage{verbatim}
\usepackage{color}

\begin{document}
%\linenumbers

\title{Unconventional spin texture driven by higher-order spin-orbit interactions}
\author{Jiaxuan Wu$^{1,3}$, Boyun Zeng$^1$, and Hanghui Chen$^{1,2}$}

\affiliation{$^1$NYU-ECNU Institute of Physics, NYU Shanghai, Shanghai 200122, China\\
$^2$Department of Physics, New York University, New York, NY 10003, USA\\
$^3$Department of Materials Science and Engineering, University of Wisconsin-Madison, Madison, WI 53706, USA}

\begin{abstract}
  Spin splitting and the resulting spin texture are central to emerging spintronic applications. In non-centrosymmetric non-magnetic materials containing heavy elements, spin textures are typically governed by low-order, momentum-dependent spin-orbit interactions, such as Rashba spin-orbit interaction with linear or cubic order in crystal momentum. In this work, we use \textit{ab initio} calculations to reveal a previously unidentified spin texture in the conduction bands of a prototypical ferroelectric nitride LaWN$_3$. In addition to the usual $\Gamma$-centered vortex, we find six new vortices and anti-vortices located at non-high-symmetry points near the Brillouin zone center. Furthermore, by combining group-theoretical analysis and $\textbf{k}\cdot\textbf{p}$ perturbation modeling, we show that, constrained by the $C_{3v}$ point group to which ferroelectric LaWN$_3$ belongs, a 7th-order Weyl spin-orbit interaction is essential to reproduce the unconventional spin structure observed in first-principles calculations. We also find that weak electron doping of LaWN$_3$ leads to a Fermi surface whose spin-arrow contour exhibits an unusual epicycloid pattern--a distinctive signature that is experimentally accessible. Our work demonstrates that higher-order spin-orbit interactions are more than perturbative corrections. They can play a dominant role in shaping the spin texture of non-centrosymmetric materials. Our results open up new avenues for designing spintronic devices that exploit multi-chiral spin textures beyond the conventional spin-orbit paradigm.
\end{abstract}

\maketitle

Spin splitting and the associated spin texture are of great importance in both fundamental sciences and applied research. In non-magnetic systems, the combination of inversion-symmetry-breaking and heavy elements (i.e. strong atomic spin-orbit coupling) gives rise to momentum-dependent spin-orbit interactions~\cite{Manchon2015}, such as the Rashba~\cite{bychkov1984properties,Rashba1959,Rashba1960}, Weyl~\cite{Weyl} and Dresselhaus~\cite{PhysRev.100.580} types. These interactions produce characteristic tangential~\cite{adma.201203199,PhysRevX.8.021067,Ishizaka2011}, radial~\cite{PhysRevLett.124.136404,PhysRevLett.125.216402,srep12024}, and tangential-radial-mixed~\cite{Appl.Phys.Lett.101.242103,pssr.201900684,Varotto2022} spin textures, respectively. Due to time reversal symmetry, all these spin-orbit interactions have an odd-order polynomial dependence on crystal momentum \textbf{k}~\cite{PhysRevB.85.075404}. The simplest and most common Rashba, Weyl and Dresselhaus spin-orbit interactions are of linear order: i.e. $H_{\textrm{R}}=\alpha_{\textrm{R}}(\sigma_xk_y - \sigma_yk_x)$, $H_{\textrm{W}}=\alpha_{\textrm{W}}(\sigma_xk_x + \sigma_yk_y)$ and $H_{\textrm{D}}=\alpha_{\textrm{D}}(\sigma_xk_x - \sigma_yk_y)$~\footnote{Rotating the $k_x$-$k_y$ coordinates by 90 degrees yields $H_{\textrm{D}}=\alpha_{\textrm{D}}(\sigma_xk_y + \sigma_yk_x)$.}, where $\sigma_x$ and $\sigma_y$ are Pauli matrices~\cite{PhysRevB.104.104408}. The corresponding spin splitting and spin texture resulting from these linear spin-orbit interactions (see Supplementary Note 3) have been extensively studied theoretically and have been widely observed experimentally~\cite{PhysRev.100.580, bychkov1984properties, Bihlmayer2022, SHANAVAS2015121, PhysRevLett.108.206601, PhysRevB.94.165202, PhysRevB.95.245141, Tao2018, adma.201203199, Djani2019, PhysRevB.93.245159, PhysRevLett.122.116401, PhysRevB.90.161108, PhysRevB.100.174415, PhysRevMaterials.3.084416, Stroppa2014, Ishizaka2011, PhysRevB.101.014109, Koralek2009, Manchon2015, acs.jpclett.0c00543, acs.jpclett.1c02596}. The pertinent material systems are very diverse, ranging from bulk ferroelectric materials like GeTe~\cite{adma.201203199} and SnTe~\cite{PhysRevB.90.161108}, to artificial heterostructures such as LaAlO$_3$/SrTiO$_3$~\cite{PhysRevLett.104.126803,PhysRevLett.104.126802,PhysRevLett.108.206601} and LaVO$_3$/KTaO$_3$~\cite{j.physe.2022.115394} interfaces. More recently, the next order cubic Rashba and Dresselhaus spin-orbit interactions have attracted increasing attention. It has been shown that the interplay of linear and cubic Rashba/Dresselhaus spin-orbit interactions can induce intricate spin splitting and complex spin texture \cite{PhysRevB.95.245141, Tao2018, adma.201203199, Djani2019, PhysRevB.93.245159, PhysRevLett.122.116401, PhysRevB.90.161108, PhysRevB.100.174415, PhysRevMaterials.3.084416, Stroppa2014, Ishizaka2011, PhysRevB.101.014109, acs.jpclett.0c00543, acs.jpclett.1c02596, Adv.Mater.2024.2313297, Tao2021, Mater.Adv.2022.3.4170}. However, beyond-cubic higher-order spin-orbit interactions are usually regarded as perturbative corrections and therefore have received limited attention. It remains an open question whether such higher-order spin-orbit interactions may play any essential role in the spin texture of real materials and if they do, what are the characteristic features in spin texture that arise from those spin-orbit interactions.

In this work, we combine first-principles calculations, group-theoretical analysis~\cite{Dresselhaus2008group} and $\textbf{k}\cdot\textbf{p}$ perturbation theory~\cite{Voon2009kp} to show that the conduction bands of a prototypical ferroelectric nitride LaWN$_3$~\cite{acs.chemmater.5b02026, J.Mater.Chem.C.2016.4.3157-3167, PhysRevB.95.014111, PhysRevMaterials.7.084411} exhibit an unconventional spin texture: in addition to the previously reported $\Gamma$-centered spin vortex, six new vortices and anti-vortices appear near the $\Gamma$ point. Furthermore, these non-trivial spin textures are centered at a non-high-symmetry point and exhibit both tangential (Rashba-type) and radial (Weyl-type) components. Group analysis finds that constrained by the $C_{3v}$ point group to which ferroelectric LaWN$_3$ belongs~\cite{PhysRevB.95.014111}, the lowest order Weyl spin-orbit interaction is of 7th-order. This 7th-order Weyl spin-orbit interaction is indispensable for fully explaining the existence of the non-$\Gamma$-centered vortices and anti-vortices, in addition to the single vortex at the $\Gamma$ point in the conduction bands of LaWN$_3$. We also show that with a weak electron doping of LaWN$_3$, the Fermi level is shifted to the relevant conduction bands. Thus the contour of spin arrows on the Fermi surface forms an epicycloid curve with six ``knots'', which can be observed via experimental techniques such as spin-resolved angle-resolved photoemission spectroscopy. Our work demonstrates that beyond-cubic higher-order spin-orbit interactions are more than perturbative corrections. They can play a dominant role in inducing nontrivial spin texture in real non-centrosymmetric materials, which may find potential applications in spintronics beyond the conventional spin-orbit paradigm. The computational details are found in the Supplementary Information.

\begin{figure}[t]
\includegraphics[width=0.75\textwidth]{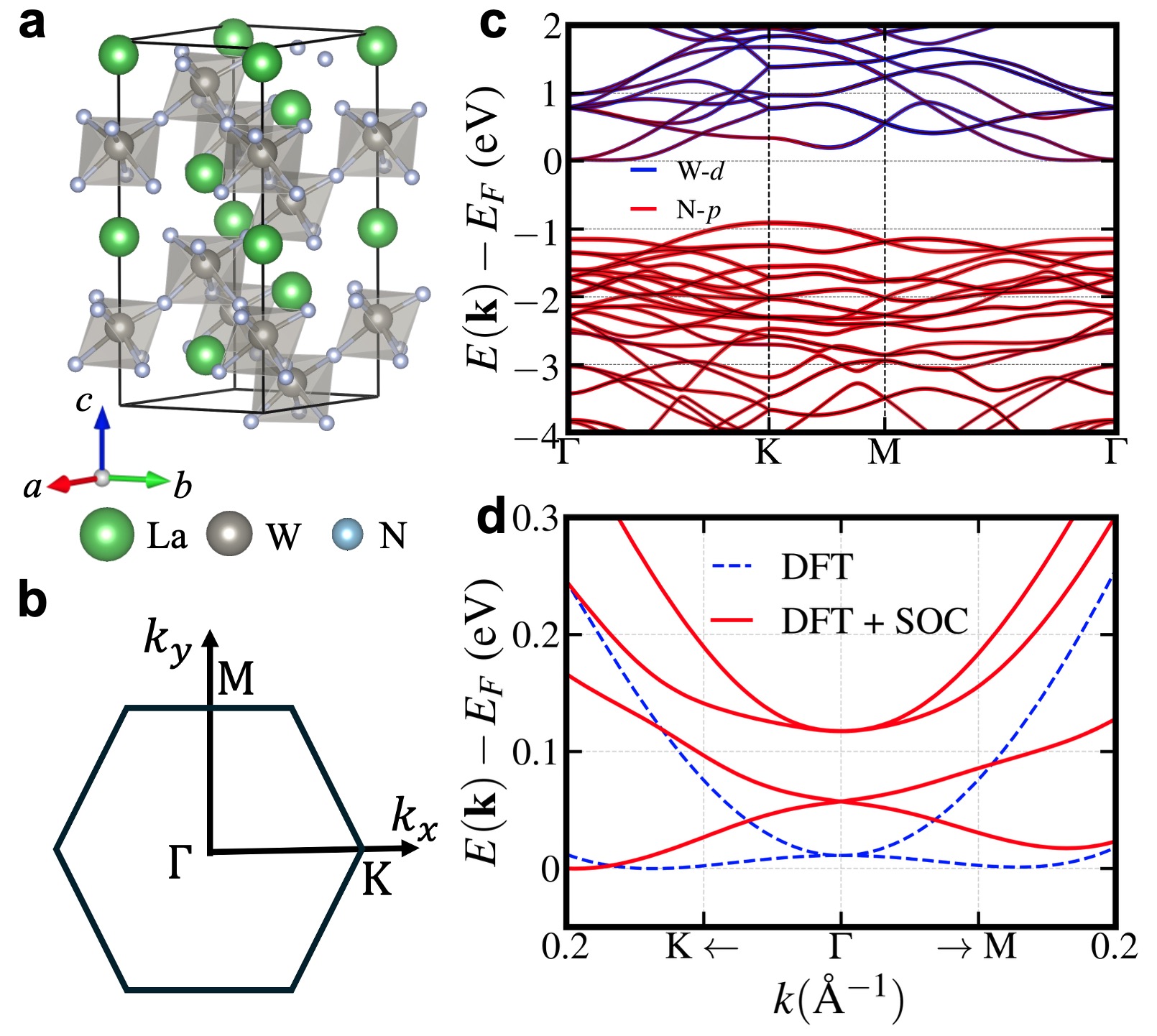}
\caption{\label{fig:fig1} (a) Crystal structure of ferroelectric LaWN$_3$ with space group $R3c$ in a hexagonal conventional cell. The green, gray and blue balls represent La, W and N atoms, respectively. All the W atoms move along the $c$ axis in each WN$_6$ octahedra, leading to a polarization. (b) The two-dimensional $k_z=0$ Brillouin zone of LaWN$_3$ in the hexagonal setting. The high-symmetry \textbf{k} points $\Gamma$, M and K are explicitly shown. (c) Electronic band structure of hexagonal LaWN$_3$ along the high-symmetry \textbf{k} path. The conduction bands are mainly composed of W-$d$ states (blue), while the valence bands dominantly consist of N-$p$ states (red). (d) The zoom-in of the conduction band edge of hexagonal LaWN$_3$. The blue dashed curves are the bands calculated without spin-orbit coupling (SOC). The red solid curves are the bands calculated with SOC.}
\end{figure}

\textit{Electronic structure}---Figure~\ref{fig:fig1}(a) shows the crystal structure of LaWN$_3$ in the hexagonal conventional cell. Bulk LaWN$_3$ is ferroelectric and crystallizes in a polar structure (space group $R3c$) with a spontaneous polarization pointing along the $c$ axis, arising from the collective off-center displacement of W atoms in the WN$_6$ octahedra~\cite{PhysRevB.95.014111, PhysRevMaterials.7.084411}. The first Brillouin zone of the $R3c$ conventional cell is shown in Fig.~\ref{fig:fig1}(b). To study the spin texture, we focus on the $k_z=0$ plane of the Brillouin zone for clarity~\cite{PhysRevB.102.041203}, in which the high-symmetry points $\Gamma$, $K$ and $M$ are explicitly shown. Fig.~\ref{fig:fig1}(c) shows the DFT-calculated band structure of $R3c$ LaWN$_3$ in a large energy window. We find that LaWN$_3$ is an insulator with a DFT band gap of about 0.8 eV, consistent with previous studies~\cite{PhysRevB.102.041203, PhysRevB.101.014109}. The atomic projections indicate that the conduction band edge is primarily composed of W-$d$ states (blue) and the valence band edge is mainly composed of N-$p$ states (red). For convenience, we shift the Fermi level to the conduction band minimum, which lies near the $\Gamma$ point. In Fig.~\ref{fig:fig1}(d), we zoom in on the band structure close to the conduction band edge and study the effects from spin-orbit coupling (SOC). The blue dashed lines are the DFT-calculated band structure, while the red solid lines are the band structure calculated by the DFT+SOC method. Without SOC, the lowest conduction bands at the $\Gamma$ point are two-fold degenerate (neglecting spin), as they belong to the $\Gamma_3$ representation of the $C_{3v}$ point group. This degeneracy can be described by using a two-orbital basis~\cite{PhysRevB.102.041203}. With SOC included, both orbital and spin degrees of freedom must be considered, where spin belongs to the $\Gamma_4$ representation (two-fold degeneracy). Consistent with Ref.~\cite{PhysRevB.102.041203}, we find that the four lowest conduction bands at the $\Gamma$ point exhibit two double-degeneracies, as shown in Fig.~\ref{fig:fig1}(d). This arises because for the $C_{3v}$ point group, $\Gamma_3 \otimes \Gamma_4 = \Gamma_4 \oplus \Gamma_5 \oplus \Gamma_6$ where the irreducible representations $\Gamma_5$ and $\Gamma_6$ are complex conjugates of each other. This yields the removal of four-fold degeneracy. However, although $\Gamma_5$ and $\Gamma_6$ are one-dimensional representations, time-reversal symmetry enforces an additional degeneracy at the $\Gamma$ point. Altogether, we obtain the 2+2 degeneracy for the four lowest conduction bands at the $\Gamma$-point in LaWN$_3$. In the following, we analyze the spin texture of these four conduction bands near the $\Gamma$ point.

\begin{figure}[t]
\includegraphics[width=1\textwidth]{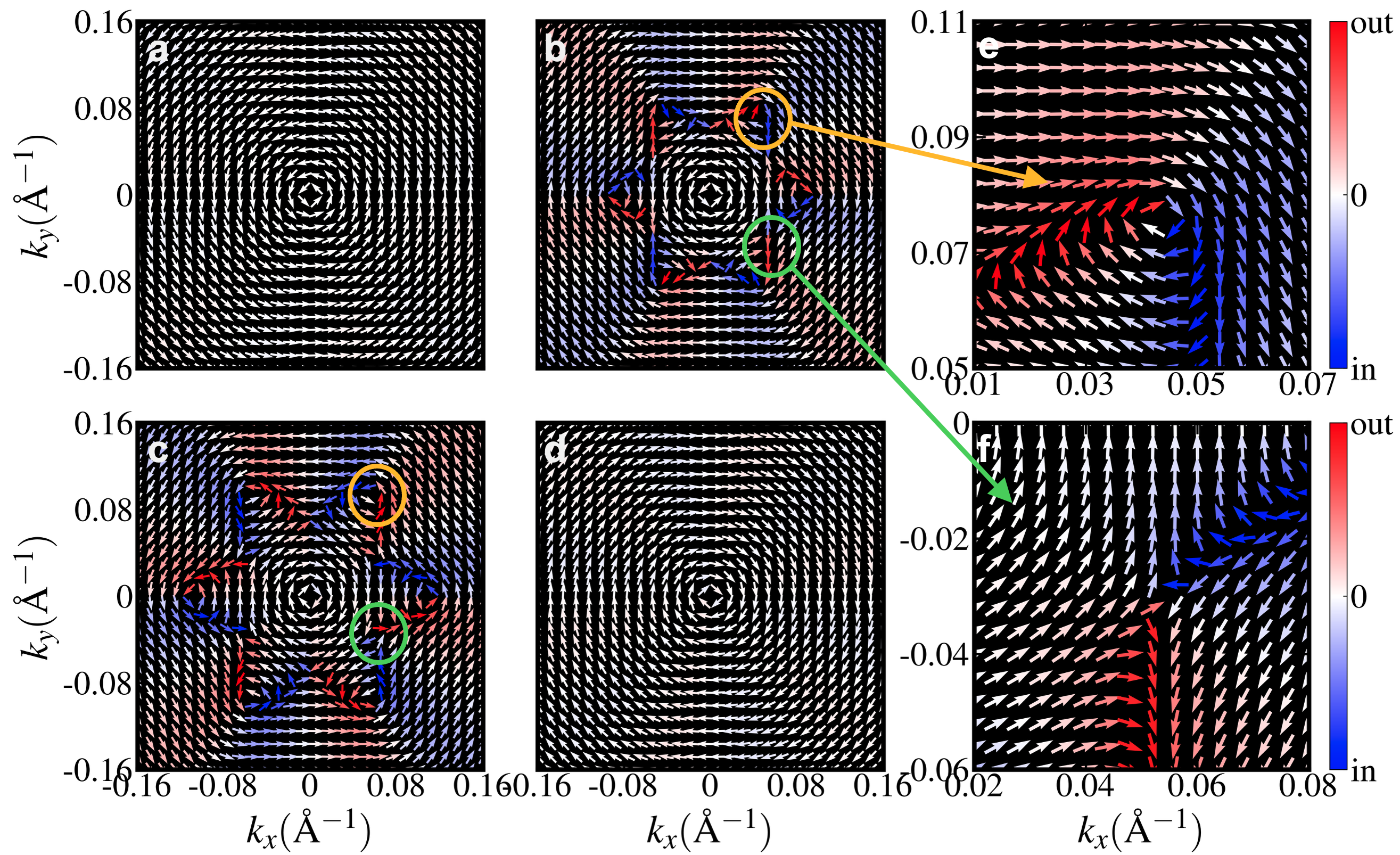}
\caption{\label{fig:fig2}Spin textures of the four lowest conduction bands of hexagonal $R3c$ LaWN$_3$ near the $\Gamma$ point in the $k_z=0$ Brillouin zone. The four lowest conduction bands are shown by the red curves in Fig.~\ref{fig:fig1}(d). (a) The lowest conduction band, in which the spin texture is a simple counter-clockwise vortex centered at the $\Gamma$ point. (b)-(c) The second and the third lowest conduction bands, in which in addition to the $\Gamma$-centered vortex, another six vortices and six anti-vortices appear, which are highlighted by an orange circle and a green circle. (d) The fourth lowest conduction band, in which the spin texture is a simple clockwise vortex centered at the $\Gamma$ point. (e)-(f) The zoom-in view of one of the vortices and the anti-vortices. The blue/red color represents the radial inward/outward spin components.}
\end{figure}

\textit{Spin texture}---%\textbf{The schematics of the linear prototypes of Rashba, Dresselhaus, and Weyl spin textures are shown in panels (a-c) in Figure~\ref{fig:fig1.5}, respectively. As Fig.~\ref{fig:fig1.5}(a) shows, the Rashba type spin texture only contains tangential components, thus forming a vortex centered at $\Gamma$ point. Fig.~\ref{fig:fig1.5}(b) shows, the Dresselhaus spin texture exhibts both tangential and radial components, forming a anti-vortex centered at $\Gamma$ point. Fig.~\ref{fig:fig1.5}(c) shows, the Weyl spin texture consists of only radial components, pointing towards/away from $\Gamma$ point, depending on the choice of electron band panel and the sign of $\alpha_W$.} 
Figure~\ref{fig:fig2} shows the spin texture of the four lowest conduction bands of LaWN$_3$ around the $\Gamma$ point. As Fig.~\ref{fig:fig1}(d) shows, for any $\textbf{k}$ point we label those four bands as 1 to 4 in ascending order of energy. For clarity, in the main text we focus on the in-plane spin texture $(s_x(\textbf{k}), s_y(\textbf{k}))$ because we find that the out-of-plane component $s_z(\textbf{k})$ follows a 60$^{\circ}$ periodic modulation and is described by a simple effective spin-orbit interaction~\cite{PhysRevLett.125.216405} (see Supplementary Note 6 for details). In Fig.~\ref{fig:fig2}, the in-plane part of each spin vector is normalized to unity and the red/blue color highlights the radial spin component that points toward/away from $\Gamma$. Fig.~\ref{fig:fig2}(a) shows that the spin texture of the 1st conduction band is a $\Gamma$-centered counter-clockwise vortex, while Fig.~\ref{fig:fig2}(d) shows that the spin texture of the 4th band is a $\Gamma$-centered clockwise vortex. Such vortices have been previously reported~\cite{PhysRevB.102.041203} and can be generated by a $\Gamma$-point Rashba spin-orbit interaction of any order. However, as shown in Fig.~\ref{fig:fig2}(b) and (c), new features appear in the spin texture of the 2nd and 3rd conduction bands. In both bands, besides the $\Gamma$-centered vortex, there are six additional vortices and anti-vortices near the $\Gamma$ point. We use an orange (green) circle to highlight one spin vortex (anti-vortex). In Fig.~\ref{fig:fig2}(e) and (f), we zoom in on the vortex and the anti-vortex around the $\Gamma$ point. The six vortices are related to each other by a rotation of 60$^{\circ}$ about the $\Gamma$ point. Between every two nearest vortices lies an anti-vortex, and thus the six anti-vortices are also related to each other by a rotation of 60$^{\circ}$ about the $\Gamma$ point. The formation of the additional vortices and anti-vortices is intriguing because their centers are neither at the $\Gamma$ point nor any other high-symmetry point. In contrast to the $\Gamma$-centered vortices shown in Fig.~\ref{fig:fig1}(a) and (d), those new vortices and anti-vortices exhibit apparent radial spin components (highlighted by the red/blue color), which cannot be accounted for purely by a $\Gamma$-point Rashba spin-orbit interaction of any order.

\begin{figure}[t]
\includegraphics[width=\textwidth]{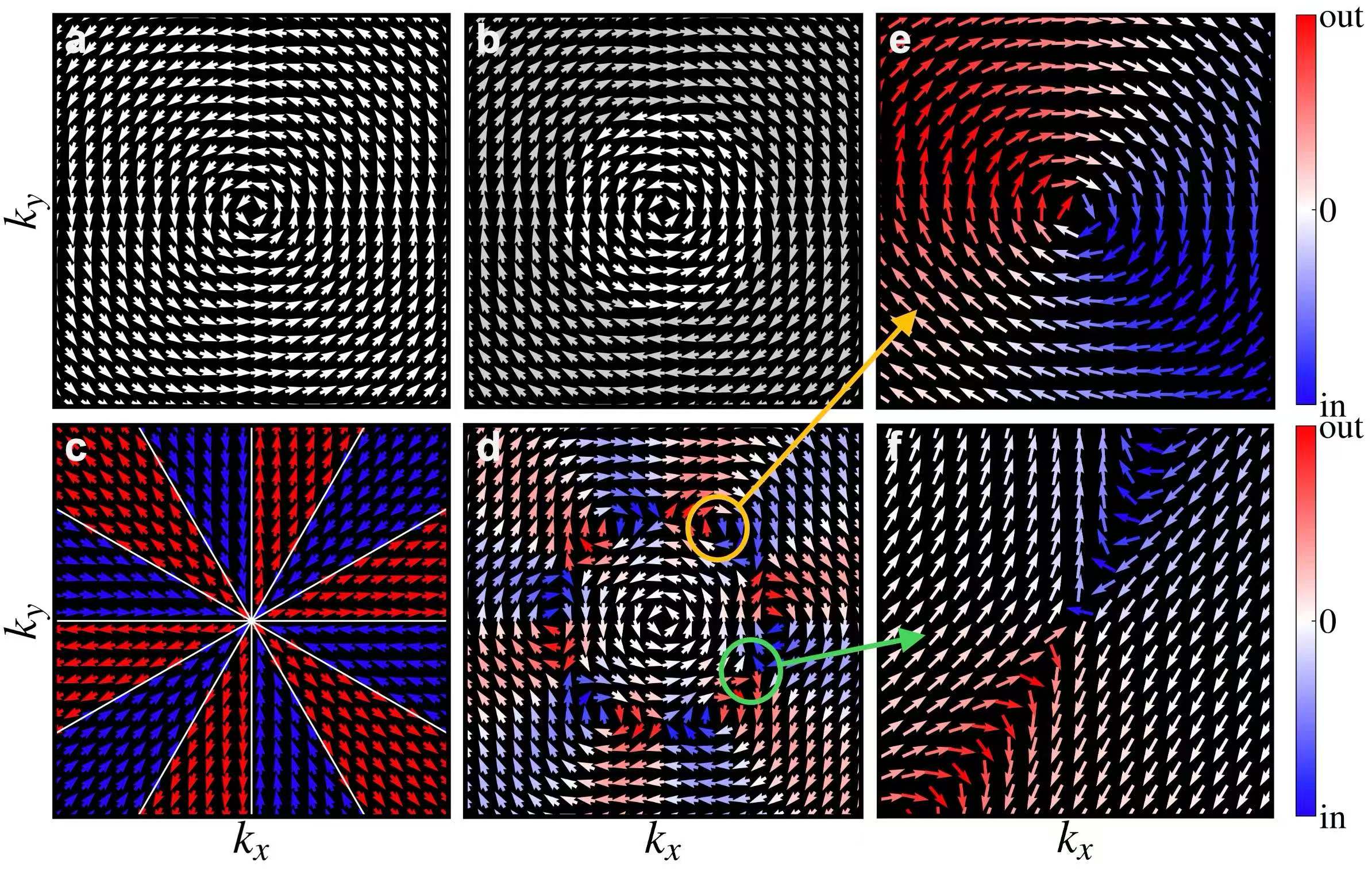}
\caption{\label{fig:fig3} Spin textures generated by different types of effective spin-orbit interactions. The blue/red color represents the radial inward/outward spin component. (a) Spin texture of a linear Rashba spin-orbit interaction. (b) Spin texture of a combination of competing Rashba spin-orbit interactions: a linear Rashba spin-orbit interaction that generates a counter-clockwise spin winding and a cubic Rashba spin-orbit interaction that generates a clockwise spin winding. (c) Spin texture of a 7th-order Weyl spin-orbit interaction. (d) Spin texture of the combination of linear+cubic Rashba spin-orbit interactions with opposite chiralities and a 7th-order Weyl spin-orbit interaction, which altogether induce additionally six vortices and six anti-vortices, besides the $\Gamma$-centered vortex. The orange and green circles highlight one vortex and one anti-vortex, which are zoomed in (e) and (f), respectively.}
\end{figure}

\textit{7th-order Weyl spin-orbit interaction}---We now analyze what $\Gamma$-point momentum-dependent spin-orbit interactions are needed to induce the aforementioned six vortices and anti-vortices found in the conduction bands of LaWN$_3$ near the Brillouin zone center. The crystal structure of $R3c$ LaWN$_3$ belongs to $C_{3v}$ point group. The large atomic number of W and the polar nature of ferroelectric LaWN$_3$ altogether give rise to effective $\Gamma$-point spin-orbit interactions. Based on the group-theoretical analysis of $C_{3v}$ point group (see Supplementary Note 4), we find that the symmetry-allowed lowest order $\Gamma$-point spin-orbit interaction is of Rashba type and is linear, taking the familiar form $k_x\sigma_y - k_y \sigma_x$. The second lowest order is also of Rashba type and is cubic in the form of $(k^2_x+k^2_y)(k_x\sigma_y - k_y \sigma_x)$. As we mentioned previously, Figure~\ref{fig:fig3}(a) shows that a $\Gamma$-point Rashba spin-orbit interaction produces a swirling spin texture, i.e. spins point along the tangent direction with a definite chirality. When multiple $\Gamma$-point Rashba spin-orbit interactions of different orders but opposite chiralities coexist, the resulting spin texture may exhibit counter-clockwise rotation for $k < k_c$ and clockwise rotation for $k>k_c$, as shown in Fig.~\ref{fig:fig3}(b). However, $\Gamma$-point Rashba spin-orbit interactions of any order or their combinations cannot produce the additional vortices and anti-vortices centered away from the high-symmetry point $\Gamma$, as observed in Fig.~\ref{fig:fig2}. These additional features require spin-orbit interactions that introduce radial spin components, such as Dresselhaus or Weyl types. Nevertheless, Dresselhaus interactions are symmetry-forbidden at the $\Gamma$ point for $C_{3v}$. On the other hand, $\Gamma$-point Weyl spin-orbit interaction is symmetry-allowed, but its lowest allowed order is 7th, in the complicated form of $k_xk_y(k_x^2-3k_y^2)(k_y^2-3k_x^2)(k_x \sigma_x + k_y \sigma_y)$ (see Supplementary Note 4). Weyl spin-orbit interaction produces a ``radiating'' spin texture~\cite{PhysRevLett.125.216405, PhysRevB.104.104408}, where spins point along the radial direction—either inward or outward. Fig.~\ref{fig:fig3}(c) illustrates the spin texture generated by the 7th-order $\Gamma$-point Weyl spin-orbit interaction allowed by $C_{3v}$ point group. All the spins point along the radial direction. For every 30$^{\circ}$, spins alternate the direction between inward and outward, as highlighted by the red and blue color, respectively. This is because the 7th-order Weyl spin-orbit interaction changes sign at $k_y = 0$, $k_y = \pm \sqrt{3} k_x$, $k_y= \pm \frac{1}{\sqrt{3}}k_x$ and $k_x = 0$. To reproduce the additional vortices and anti-vortices observed in the conduction bands of LaWN$_3$, both Rashba and Weyl spin-orbit interactions at the $\Gamma$ point are required—so that spins not only swirl but also develop radial components around non-high-symmetry centers.  Explicitly, in Fig.~\ref{fig:fig3}(d), we present the spin texture of a linear Rashba spin-orbit interaction + a cubic Rashba spin-orbit interaction + a 7th-order Weyl spin-orbit interaction, all of which are symmetry-allowed under the $C_{3v}$ point group. We find that, in addition to the vortex centered at the $\Gamma$ point, six additional vortices and anti-vortices are successfully reproduced near the $\Gamma$ point, as observed in the conduction bands of LaWN$_3$ (see Fig.~\ref{fig:fig2}(b) and (c)). Figures~\ref{fig:fig3}(e) and (f) show a close-up view of one of the vortices and one of the anti-vortices, respectively, closely resembling those in Fig.~\ref{fig:fig2}(e) and (f). Since Weyl spin-orbit interaction is indispensable for generating these non-$\Gamma$-centered vortices/anti-vortices—and the 7th-order Weyl spin-orbit interaction is the lowest-order symmetry-allowed Weyl interaction in $C_{3v}$ point group—the presence of those six additional vortices and anti-vortices in the conduction bands of LaWN$_3$ constitutes a direct manifestation of higher-order momentum-dependent spin-orbit interaction (at least 7th-order) in real non-centrosymmetric materials.

\begin{figure}[t]
\includegraphics[width=0.8\textwidth]{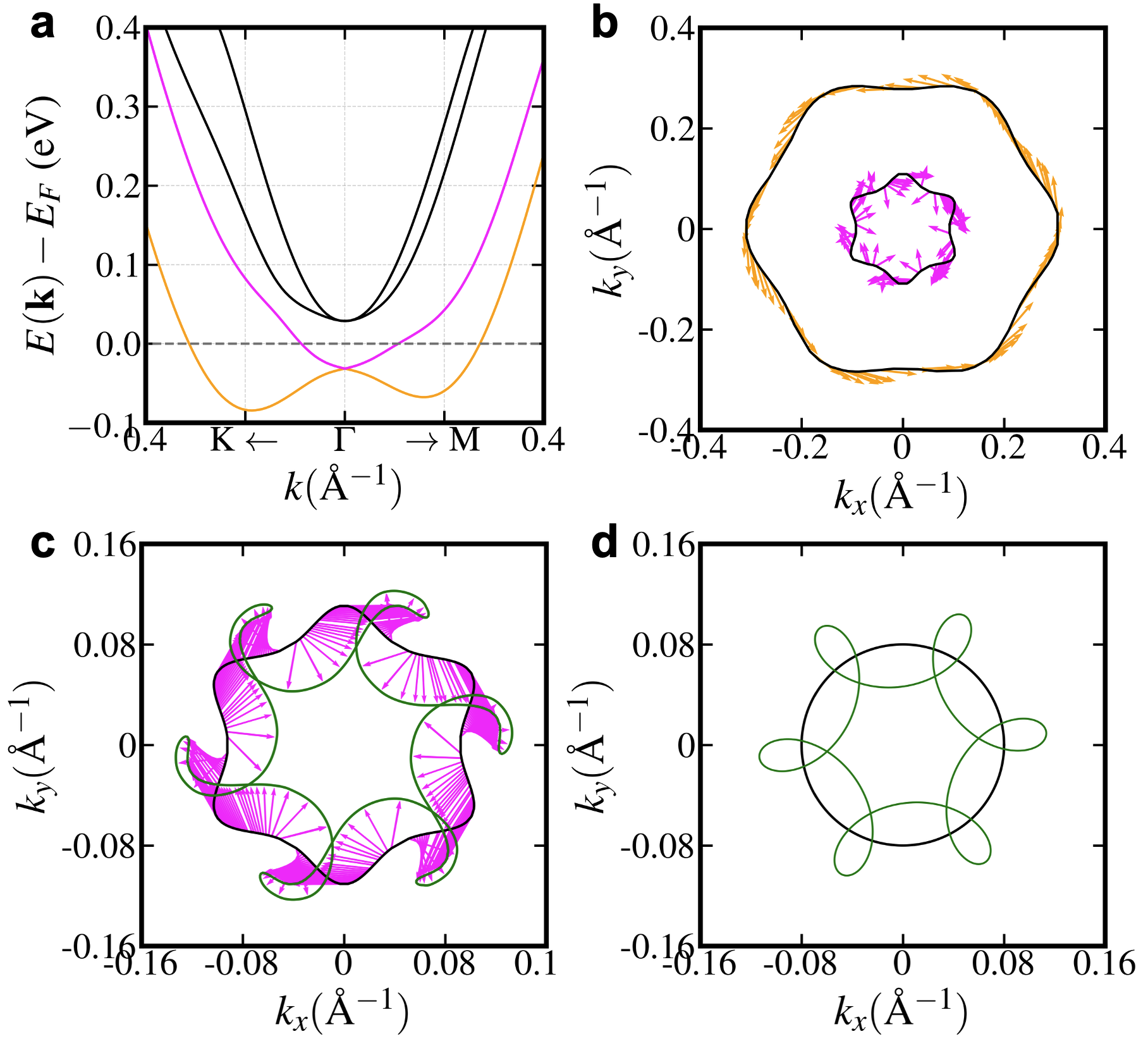}
\caption{\label{fig:fig4} (a) Band structure of electron-doped
  LaWN$_3$ with a carrier concentration of 0.017$e$ per W
  (i.e. $n_{\rm{3D}}\simeq 2.6\times 10^{20}$ cm$^{-3}$), where the
  Fermi level cuts through the two lowest conduction bands near
  the $\Gamma$ point, highlighted by the orange and pink curves.  (b)
  Two-dimensional $k_z=0$ Fermi surface of electron-doped LaWN$_3$,
  which arises from the two lowest conduction bands (the orange and
  pink curves in (a)). The spin textures of those two conduction bands
  are shown by the orange and pink arrows, respectively. (c) Zoom-in
  of the spin texture of the second lowest conduction band. The black
  curve is the Fermi surface and the green curve is the contour of the
  pink spin arrows. (d) An ideal circular Fermi surface (black curve)
  and an epicycloid curve with six ``knots'' (green curve), which are
  topologically equivalent to the Fermi surface and the contour of spin arrows
  shown in (c).}
\end{figure}

\textit{Experimental verification}---Finally, we discuss how one may experimentally access this unconventional spin texture using techniques such as spin-resolved angle-resolved photoemission spectroscopy. We study weakly electron-doped LaWN$_3$, which is achievable via slight nitrogen deficiency. Figure~\ref{fig:fig4}(a) shows the band structure of electron-doped LaWN$_3$ at a carrier concentration of $2.6\times 10^{20}$ cm$^{-3}$ (0.017$e$ per formula). At the conduction band edge, the first and second conduction bands (highlighted in orange and pink) cross the Fermi level, forming a Fermi surface. Fig.~\ref{fig:fig4}(b) shows the corresponding Fermi surface in the $k_z=0$ plane. The spin textures of the two bands on the Fermi surface are indicated explicitly by the orange and pink arrows. On the lowest conduction band (outer circle), the orange arrows exhibit a $\Gamma$-centered vortex pattern. On the second-lowest conduction band (inner circle), the pink arrows display a more complex texture due to the 7th-order Weyl spin-orbit interaction. We zoom in on the spin texture of the second conduction band in Fig.~\ref{fig:fig4}(c), where the black curve denotes the Fermi surface and the green curve traces the contour of spin directions. While both the Fermi surface and the spin contour exhibit intricate shapes, they are topologically equivalent to a circle and an epicycloid, respectively, as illustrated in Fig.~\ref{fig:fig4}(d). The epicycloid arises from the combination of two motions (see Supplementary Video \texttt{epicycloid.mp4}): the revolution of the circle around the $\Gamma$ point and its rotation about its own center. The rotation occurs six times faster than the revolution, resulting in six ``knots''. These knots are indicators of the underlying spin vortices and anti-vortices identified in Fig.~\ref{fig:fig2}.

Before concluding, we comment that the 7th-order $\Gamma$-point Weyl spin-orbit interaction is a generic feature of the $C_{3v}$ point group. Therefore, the spin texture of the six additional vortices and anti-vortices near the $\Gamma$ point can also be expected in other materials belonging to the $C_{3v}$ symmetry class. We find that the widely studied KTaO$_3$/LaAlO$_3$ (111) interface~\cite{science.aba5511,science.abb3848,sciadv.adf1414,Nat.Commun14.951,pssr.202200441,Adv.Mater.2024.2313297,Nat.Commun15.7704,pssr.202200441,APL.Mater.11.121108,aelm.201800860} also exhibits a similar unconventional spin texture, driven by the same higher-order Weyl spin-orbit interaction. Further details are provided in Supplementary Note 9.  

In conclusion, our work shows that in non-centrosymmetric non-magnetic materials, beyond-cubic higher-order momentum-dependent spin-orbit interactions are more than mere perturbative corrections. They can play an essential role in inducing unconventional spin texture. In particular, by using first-principles calculations and symmetry-based $\textbf{k}\cdot\textbf{p}$ perturbation theory, we find that constrained by the $C_{3v}$ group symmetry, a 7th-order Weyl spin-orbit interaction is indispensable for explaining the six additional vortices and anti-vortices observed in the conduction bands of $R3c$ LaWN$_3$ near the $\Gamma$ point. Unlike the usual Rashba-like chiral spin windings centered at high-symmetry points, these spin textures exhibit both tangential (Rashba-type) and radial (Weyl-type) components, and are centered at a non-high-symmetry point near $\Gamma$. Furthermore, we estimate that electron doping of LaWN$_3$ to a carrier concentration of approximately $2.6 \times 10^{20}~\mathrm{cm}^{-3}$ is sufficient to shift the Fermi level into the relevant conduction bands, making experimental observation of these spin textures feasible via techniques such as spin-resolved angle-resolved photoemission spectroscopy.     

%\begin{acknowledgments}
% We are grateful to XXX for useful discusssions. This project was financially supported by the National Natural Science Foundation of China under project number 12374064 and 12434002, Science and Technology Commission of Shanghai Municipality under grant number 23ZR1445400 and a grant from the New York University Research Catalyst Prize under project number RB627. NYU High-Performance-Computing (HPC) provides computational resources.
%\end{acknowledgments}

\newpage
\clearpage
\bibliography{main_citation}

\end{document}